\documentclass[iop,revtex4]{emulateapj}
\usepackage{amsmath}
\usepackage{threeparttable}

\slugcomment{Accepted for publication in ApJ}

\shorttitle{Meteoritic Abundances of Nucleobases and Potential Reaction Pathways}
\shortauthors{Pearce \& Pudritz}

\usepackage{etex}
\usepackage{m-pictex}
\usepackage{m-ch-en}

\begin{document}

\title{Seeding the Pregenetic Earth: Meteoritic Abundances of Nucleobases and Potential Reaction Pathways}

\author{Ben K. D. Pearce\altaffilmark{1,4} and Ralph E. Pudritz\altaffilmark{2,3,5} }

\altaffiltext{1}{Department of Physics and Astronomy, University of British Columbia, HENN 325, 6224 Agricultural Rd, Vancouver, BC, V6T 1Z1, Canada}
\altaffiltext{2}{Origins Institute, McMaster University, ABB 241, 1280 Main St, Hamilton, ON, L8S 4M1, Canada}
\altaffiltext{3}{Department of Physics and Astronomy, McMaster University, ABB 241, 1280 Main St, Hamilton, ON, L8S 4M1, Canada}
\altaffiltext{4}{E-mail: ben.pearce@alumni.ubc.ca}
\altaffiltext{5}{Email: pudritz@physics.mcmaster.ca}

\begin{abstract}

Carbonaceous chondrites are a class of meteorite known for having a high content of water and organics. In this study, abundances of the nucleobases, i.e., the building blocks of RNA and DNA, found in carbonaceous chondrites are collated from a variety of published data and compared across various meteorite classes. An extensive review of abiotic chemical reactions producing nucleobases is then performed. These reactions are then reduced to a list of 15 individual reaction pathways that could potentially occur within meteorite parent bodies. The nucleobases guanine, adenine and uracil are found in carbonaceous chondrites in the amounts of 1--500 ppb. It is currently unknown which reaction is responsible for their synthesis within the meteorite parent bodies. One class of carbonaceous meteorites dominate the abundances of both amino acids and nucleobases---the so-called CM2 (e.g. Murchison meteorite). CR2 meteorites (e.g. Graves Nunataks) also dominate the abundances of amino acids, but are the least abundant in nucleobases. The abundances of total nucleobases in these two classes are $330 \pm250$ ppb and $16 \pm13$ ppb respectively. Guanine most often has the greatest abundances in carbonaceous chondrites with respect to the other nucleobases, but is 1$\textendash$2 orders of magnitude less abundant in CM2 meteorites than glycine (the most abundant amino acid). Our survey of the reaction mechanisms for nucleobase formation suggests that Fischer-Tropsch synthesis (i.e. CO, H$_2$ and NH$_3$ gases reacting in the presence of a catalyst such as alumina or silica) is the most likely for conditions that characterize early states of planetesimals.


\end{abstract}

\section{Introduction}

It has long been proposed that a significant portion of the prebiotic reservoir of biomolecules (including amino acids, fatty acids and nucleobases) on Earth was synthesized in comets and planetesimals, and delivered by meteorites, comets and interplanetary dust particles \citep{ChamberlinChamberlin,Oro1961a,Chyba1992,PierazzoChyba,Burton2012}. (Planetesimals in this context are solid bodies approximately 1 km to 100 km in diameter, originating from the circumstellar disk.) This theory is substantiated by the high abundances of organics measured in the fragments of planetesimals, i.e., the carbonaceous chondrite meteorites that have fallen on the Earth. Additional support comes from the recent analysis of biomolecules on the surface of Comet 67P/Churyumov-Gerasimenko, which has revealed some evidence for nonvolatile organic macromolecular materials \citep{Capaccioni}. Thus, the formation and origin of pregenetic biomolecules for young planets is of particular interest in understanding the steps that led to the appearance of life.

Nucleobases are the nitrogen-containing, characteristic molecules in nucleotides. They come in two varieties: the purines, guanine (G) and adenine (A), and the pyrimidines, cytosine (C), thymine (T) and uracil (U). They are important because the particular ordering of nucleotides in the chain(s) of DNA in organisms (or RNA in some viruses) constitutes the genome. In DNA the base pairs are G-C and A-T, and in RNA the base pairs are G-C and A-U. Since all known organisms and viruses use DNA and/or RNA to replicate, and since DNA or RNA stores the instructions for building proteins, understanding the origin of the nucleobases is essential to understanding the origin of the genetic code.

As part of a long term project to understand the astrophysical origin of biomolecules and their delivery to forming planets, \citet{Cobb2014} first collated and displayed the abundances of amino acids in carbonaceous chondrites.  Theoretical work on the origin of amino acids by means of aqueous Strecker reactions occurring in meteorite parent bodies was then carried out and compared with the meteoritic record \citep{CobbPudritzPearce}. In this and subsequent papers, we extend this approach to nucleobases by first presenting the available data on nucleobase abundances within meteorites.

There are three components to the nucleotide: one nucleobase (G, A, C, T or U), one sugar (ribose or deoxyribose) and at least one phosphate group (PO4$^{3-}$). Three of the five nucleobases in DNA or RNA (G, A and U) have been discovered in CM2, CR2, CI1, CM1, CR1 and CR3 type meteorites \citep{Callahan2011,Martins2008,Shimoyama1990,StoksSchwartz1981,StoksSchwartz1979,vanderVeldenSchwartz,Hayatsu1975,Hayatsu1968,Hayatsu1964}, along with some nucleobase analogs (purine, 2,6-diaminopurine, 6,8-diaminopurine) and nucleobase catabolic intermediates (hypoxanthine and xanthine). Ribose and deoxyribose have not been identified in meteorites (although the sugar dihydroxyacetone has \citep{Cooper2001}). Phosphates have been found as minerals in many meteorites, typically as Ca-phosphates or Fe-phosphate \citep{EbiharaHonda}. For informative reviews about carbonaceous chondrite meteorites and their classification, please see \citet{Cobb2014}, \citet{Weisberg}, \citet{BottaBada} and \citet{Hayes1967}.

We begin in this paper by compiling a comprehensive list of the observed abundances of nucleobases in carbonaceous chondrites as reported in the scientific literature. We then collate and display the data from these studies, emphasizing the trends in abundances and frequencies of these molecules by meteoritic type---and hence formation conditions. The data is first displayed by individual nucleobase abundance for each meteorite sample, then by total nucleobase abundance for each meteorite sample, and finally by average relative nucleobase abundance to G for CM2 meteorites. The potential for nucleobase contamination in the samples of each study are also reviewed.

We then perform an extensive survey of all of the published chemical methods that have been employed or suggested as pathways for the abiotic formation of nucleobases. This survey is presented as a starting point in order to understand the reaction pathway(s) that could occur within planetesimals. The comprehensive list is reduced by disregarding reactions that are unlikely to occur in such environments. A final list of candidate nucleobase reaction pathways within planetesimals is then proposed.

In a subsequent paper, we will use this candidate list to investigate the synthesis of nucleobases in planetesimal interiors.

\section{Meteoritic Data}

The two fundamental techniques for measuring meteoritic nucleobases in the laboratory are chromatography and mass spectrometry. Chromatography separates a mixture by dissolving it in a fluid, a.k.a. the mobile phase, and sending it through a structure, e.g., a column, holding a stationary phase, e.g., a silica layer. The molecules in the mixture travel at different speeds through the structure, causing them to separate for detection. Mass spectrometry separates a mixture by ionizing its constituents (e.g. by bombarding them with electrons) and then accelerating them through an electric or magnetic field. The ions are deflected as they pass through the field, and separated according to their mass-to-charge ratio. 

The specific analytical techniques used to detect nucleobases in carbonaceous chondrites have improved significantly over time. First analyses of nucleobases applied Paper and Thin-layer Chromatography (PC, TLC) \citep{Hayatsu1968,Hayatsu1964} and Mass Spectrometry (MS) \citep{Hayatsu1975}. (PC and TLC are very similar, with the only difference being that PC has its stationary phase within a tube and TLC has its stationary phase on a plane.) These studies published extraordinarily high concentrations of G and A in the Murchison and Orgueil meteorites, with abundances ranging from 5000--20000 ppb. Analytical techniques then moved towards Ion-exclusion Chromatography (IEC, IEC-MS) \citep{StoksSchwartz1981,StoksSchwartz1979,vanderVeldenSchwartz} and High Performance Liquid Chromatography (HPLC, HPLC-MS) \citep{Callahan2011,Shimoyama1990}. These studies yielded nucleobase abundances in the 1--500 ppb range. IEC uses an aqueous mobile phase and separates the ionized compounds from the non-ionized compounds of a mixture by excluding the former. After the mixture is separated by the chromatograph, it is sometimes sent into mass spectrometer for further separation. HPLC has a liquid mobile phase and improves chromatographic resolution by applying pressure to the dissolved mixture, allowing the mixture to flow at a much higher speed through the column. The use of a higher pressure allows for smaller particles to be used for the column packing material, which creates a greater surface area for interaction between the stationary phase and the flowing molecules. This allows for a better separation of molecules. MS is also sometimes used in tandem with HPLC for further separation.

TLC has been found to be less sensitive and gives worse separation than HPLC \citep{Choma2005}. \citet{StoksSchwartz1981} also note that the analytical techniques applied in the earliest studies (PC, MS) are not very suitable for quantitative purposes. Therefore due to their lower accuracy and sensitivity, we won't consider the studies that used PC, TLC or MS in the quantitative analysis in this paper (i.e. \citet{Hayatsu1975,Hayatsu1968} and \citet{Hayatsu1964}).

\begin{deluxetable*}{lllll}
\tablecolumns{5}
\tablecaption{Meteoritic nucleobase data sources\label{DataSources}.}
 \tablehead{Type&Meteorite&Sample Number&Analytical Technique&Reference(s)}
 \startdata
CI1 & Orgueil & Ehrenfreund & HPLC-MS & \citet{Callahan2011}\\
CI1 & Orgueil & MNHN & IEC-MS & \citet{StoksSchwartz1981}; \citet{StoksSchwartz1979}\\
CM1 & Scott Glacier & SCO 06043 & HPLC-MS & \citet{Callahan2011}\\
CM1 & Meteorite Hills & MET 01070 & HPLC-MS & \citet{Callahan2011}\\
CM2 & Allan Hills & ALH 83100 & HPLC-MS & \citet{Callahan2011}\\
CM2 & Lewis Cliff & LEW 90500 & HPLC-MS & \citet{Callahan2011}\\
CM2 & Lonewolf Nunataks & LON 94102 & HPLC-MS & \citet{Callahan2011}\\
CM2 & Murchison & Smithsonian & HPLC-MS & \citet{Callahan2011}\\
CM2 & Yamato & Yamato 74662 & HPLC & \citet{Shimoyama1990}\\
CM2 & Yamato & Yamato 791198 & HPLC & \citet{Shimoyama1990}\\
CM2 & Murchison & ASU 1 & IEC-MS & \citet{StoksSchwartz1981}; \citet{StoksSchwartz1979}\\ 
CM2 & Murchison & ASU 2 & IEC & \citet{vanderVeldenSchwartz}\\
CM2 & Murray & N.A. & IEC-MS & \citet{StoksSchwartz1981}; \citet{StoksSchwartz1979}\\
CR1 & Grosvenor Mountains & GRO 95577 & HPLC-MS & \citet{Callahan2011}\\
CR2 & Elephant Moraine & EET 92042 & HPLC-MS & \citet{Callahan2011}\\
CR2 & Graves Nunataks & GRA 95229 & HPLC-MS & \citet{Callahan2011}\\
CR3 & Queen Alexandra Range & QUE 99177 & HPLC-MS & \citet{Callahan2011}\\[-3mm]
 \enddata
\end{deluxetable*}

Table~\ref{DataSources} lists the data sources for the nucleobases, nucleobase intemediates and nucleobase analogs in each carbonaceous chondrite in this analysis. Carbonaceous chondrites are classified into types based primarily on composition and secondarily on amounts of aqueous and thermal processing (see \citet{Cobb2014} for an overview). For instance, the character 'R' in CR1 meteorites signifies the meteorite is of Renazzo-like composition. The Renazzo meteorite is a meteorite that fell in Renazzo, Italy in 1824. The number '1' in CR1 meteorites signifies the meteorite is of petrologic type 1 and therefore underwent the highest amount of aqueous alteration. The level of aqueous alteration is determined by the amount the chondrules (round grains) in the meteorite have been changed by the presence of water. Petrologic type 1 signifies the complete obliteration of chondrules in the meteorite, whereas petrologic type 3 signifies the chondrules in the meteorite have not been altered. The majority of meteorite samples with reported nucleobase abundances are of the CM2 type (Mighei-like composition with an average level of aqueous alteration).

\subsection{Contamination Possibility}

There is always a concern about terrestrial contamination when analyzing meteorites for organics. \citet{Callahan2011} addressed this by analyzing the content of the terrestrial environment around which the meteorites were retrieved. In the case of Antarctic meteorites they analyzed dried residue from an ice sample collected in 2006 from the Graves Nunataks region, and in the case of the Murchison meteorite they analyzed a soil sample collected in 1999 from near the original Murchison meteorite fall site. In the Antarctic residue sample, less than 5 pptr (parts-per-trillion) of G and A were measured, strongly suggesting the meteorite nucleobases to be of extraterrestrial origin. Interestingly 1380 ppb of C was measured in the soil sample from near the Murchison fall site, but no abundance of C was measured in the Murchison meteorite sample. Also, several nucleobase analogs were measured in the Murchison meteorite that are either rare or absent on Earth (purine, 2,6-diaminopurine, and 6,8-diaminopurine). The conjunction of these findings strongly suggests an extraterrestrial origin of these meteorite nucleobases. 

\citet{Shimoyama1990} argued that because the amino acid mixtures measured in the analyzed meteorite were racemic, i.e., composed of equal amounts of D and L enantiomers, it is unlikely that the sample (also containing nucleobases) was subjected to much contamination. Other studies took the high uracil-to-thymine ratios in the sample (e.g. 21 in a sample of the Murchison meteorite) as evidence that the nucleobases are extraterrestrial in origin \citep{StoksSchwartz1981,StoksSchwartz1979,vanderVeldenSchwartz}. A high uracil-to-thymine ratio is suggestive of an extraterrestrial origin of nucleobases because studies on biogenetic nucleobases in various soils and recent sediments have measured little uracil with respect to the other nucleobases \citep{StoksSchwartz1979}.

\section{Results: Nucleobase Abundances and Relative Frequencies}

\subsection{Nucleobase Abundances in Meteorites} 

\begin{figure*}[hbtp]
\centering
\includegraphics[width=\textwidth]{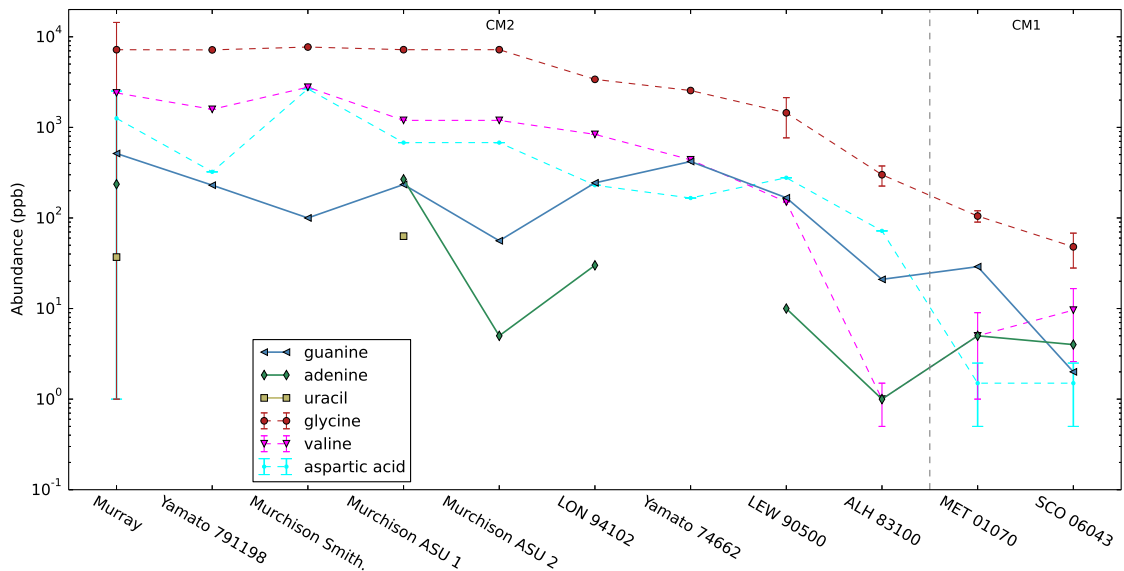}
\caption{Abundances of three nucleobases (G: blue, A: green, U: yellow) and three amino acids (glycine: red, valine: magenta, aspartic acid: cyan) in CM meteorites. The meteorite order is monotonically decreasing with respect to glycine. Amino acid data is obtained from \citet{Cobb2014} and \citet{Shimoyama1979}.}
\label{CMAbundances}
\end{figure*}

We display the individual nucleobase abundance data grouped by carbonaceous chondrite type. The CM meteorites are displayed in one plot, and the CI and CR meteorites are grouped together in another plot---due to limited nucleobase data of the latter group. The ordering of meteorite samples for both plots is the same as the ordering defined by \citet{Cobb2014} for amino acid abundances. This order was chosen by \citet{Cobb2014} to give a smooth monotonic sequence for the most abundant amino acid: glycine. In Figure~\ref{CMAbundances}, we illustrate the ppb (parts-per-billion) abundances of the purines G and A and the pyrimidine U in CM meteorites compared with the abundances of glycine, valine and aspartic acid found in the same meteorites from \citet{Cobb2014} and (in the case of Yamato 74662) \citet{Shimoyama1979}. G is typically the most abundant nucleobase in these samples, ranging from 2--515ppb. These G abundances are approximately 1--2 orders of magnitude lower than the glycine measured in the same meteorites. The range of G abundances in CM meteorites is closer to the intermediate and less abundant amino acids in carbonaceous chondrites: valine and aspartic acid. G is approximately 1 order of magnitude lower to 1 order of magnitude higher than both the valine and aspartic acid measured in the same meteorites. G and aspartic acid remain in relatively close abundance with each other. A and U are generally less abundant than G by a factor of 2--10. 

\begin{figure}[hbtp]
\centering
\includegraphics[width=80mm]{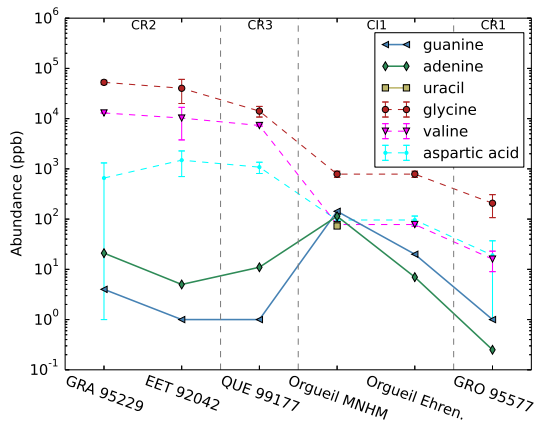}
\caption{Abundances of three nucleobases (G: blue, A: green, U: yellow) and three amino acids (glycine: red, valine: magenta, aspartic acid: cyan) in CR and CI meteorites. The meteorite order is monotonically decreasing with respect to glycine. Amino acid data is obtained from \citet{Cobb2014}.}
\label{CRIAbundances}
\end{figure}

The G abundance curve in Figure~\ref{CMAbundances} has a localized peak in the middle, but still trends downwards overall, generally following the glycine curve. The A abundance curve in Figure~\ref{CMAbundances} trends downward as well, and appears to have a local maximum in the same location the G curve did. There is not enough U data to say anything about its curve.
The correspondence between the amino acid and nucleobase curves suggests that conditions that favour the synthesis of glycine are related to those that optimize the production of nucleobases. The 1--2 orders of magnitude difference in abundance between nucleobases and glycine could have several causes: different decay rates, difference of synthesis or competing chemical reactions for the same basic materials. These will be future considerations of our theoretical work. The similarity in the bumps created by the G and A curves in Figure~\ref{CMAbundances} could suggest a similar mechanism is used to synthesize G and A within the meteorite parent bodies.

The nucleobase abundances in CI and CR meteorites are displayed in Figure~\ref{CRIAbundances}; again, along with the abundances of glycine, valine and aspartic acid from \citet{Cobb2014} as a guide. These abundances are much more reduced than in the CM meteorites, most laying in the 1-20 ppb range. The Orgueil meteorite obtained from the Mus\'{e}um National d'Histoire Naturelle (MNHN) is the only exception with slightly higher abundances of nucleobases than the other samples ($\sim$100 ppb). Abundances of nucleobases in CI and CR meteorites range from 1--4 orders of magnitude lower than glycine abundances in the same meteorites. Nucleobases range up to 4 orders of magnitude lower than valine, and up to 3 orders of magnitude lower than aspartic acid in CI and CR meteorites.

Due to the relatively low abundances of nucleobases in CM1, CR1, CR3 and CI1 meteorites in comparison to CM2 meteorites, and from the relatively low abundances of amino acids in the same meteorite types in \citet{Cobb2014}, it is likely that the parent bodies of CM1, CR1, CR3 and CI1 meteorites have a less suitable environment for organic synthesis than the parent bodies of CM2 meteorites.

\begin{figure*}[hbtp]
\centering
\includegraphics[width=\textwidth]{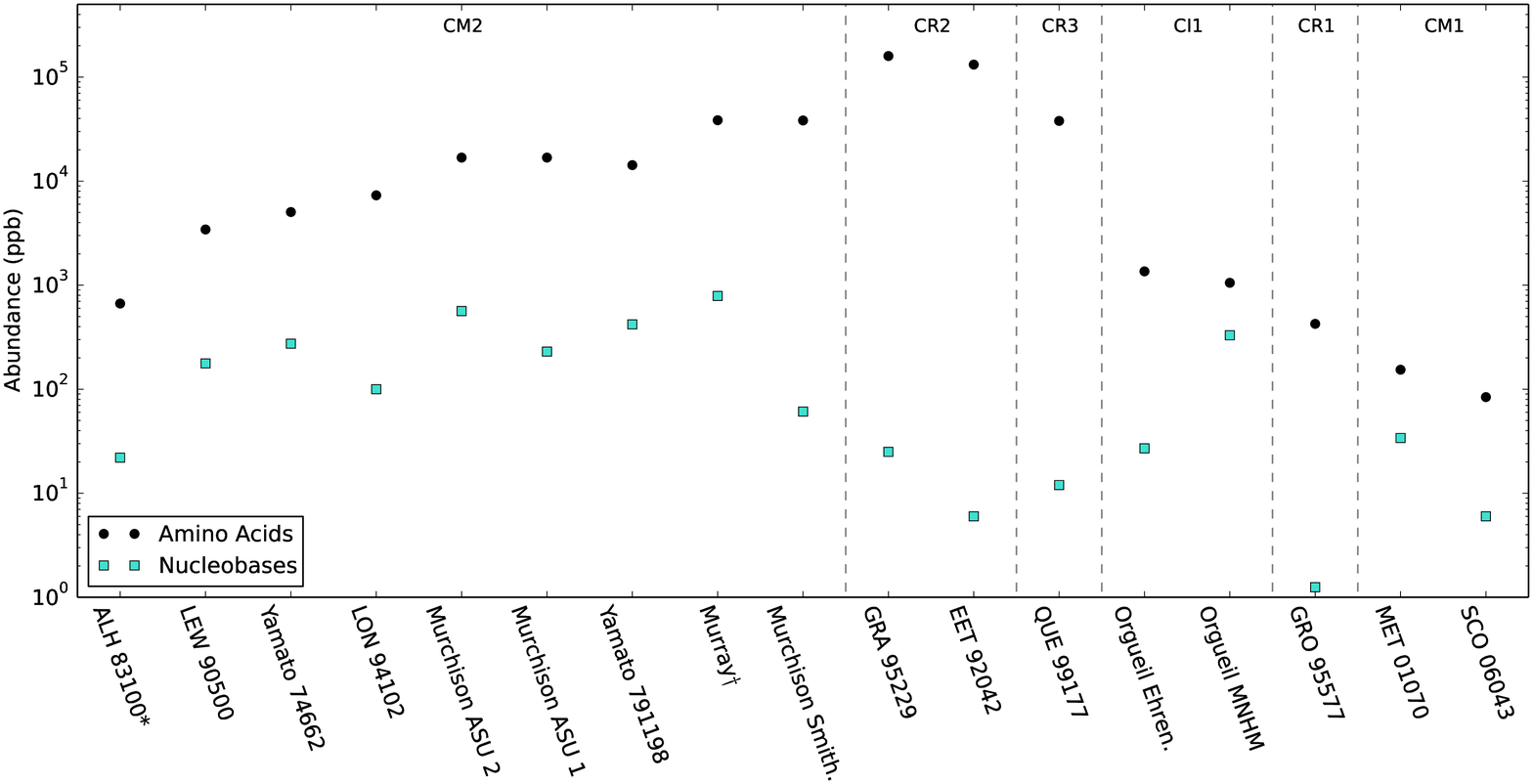}
\caption{Total nucleobase abundances from 17 meteorite samples plotted with the total amino acid abundances found in the same meteorites from \citet{Cobb2014} and \citet{Shimoyama1979}. The amino acid data taken from \citet{Cobb2014} is slightly refined. Nucleobase sum includes G, A and U. Amino acid sum includes glycine, alanine, serine, aspartic acid, glutamic acid and valine. The meteorite order conforms to the ordering of the total amino acid abundance plot in \citet{Cobb2014}. *ALH 83100 is a CM1/2 type meteorite. $\dag$The total amino acid abundance is the "Murray 1" abundance from \citet{Cobb2014}.}
\label{TotalAbundances}
\end{figure*}

Other purines were also measured in CM, CI and CR meteorites. Xanthine, a catabolic intermediate of G, was found in abundances of 4--2300 ppb and hypoxanthine, a catabolic intermediate of A, was found in abundances of 3--243 ppb. Nucleobase analogs purine, 2,6-diaminopurine, and 6,8-diaminopurine were also measured in lesser abundances (0.2--9 ppb). 

In Figure~\ref{TotalAbundances}, we sum up G, A and U in all the meteorite sources and plot the total nucleobase abundances by meteorite type. The total amino acid abundances measured in the same meteorites in this study are added to this plot for comparison. The total amino acid abundances include glycine, alanine, serine, aspartic acid, glutamic acid and valine, and are taken from \citet{Cobb2014} and \citet{Shimoyama1979}. The total amino acid abundances from \citet{Cobb2014} are verified from the original sources, and in some cases refinements are made. 

The ordering of meteorites is in accordance with the total amino acid abundance plot in \citet{Cobb2014}. It is clear from this plot that CM2 meteorites dominate in total nucleobase abundance, averaging at $330 \pm250$ ppb. This is interesting because CM2 meteorites also appear to be one of the most abundant in amino acids, with total abundances averaging $\sim$10$^4$ ppb. This could suggest a similarity in the synthesis mechanisms of amino acids and nucleobases in the parent bodies of CM2 meteorites, e.g., both sharing HCN as a reactant. Conversely CR2 meteorites, which appear to be the most abundant in total amino acids (averaging $\sim$10$^5$ ppb), are one of the least abundant in nucleobases, with a total average abundance of $16 \pm13$ ppb. These differences could arise from different reaction rates, decay rates or chemical composition within the CR2 parent bodies.

The total abundances of nucleobases in Figure~\ref{TotalAbundances} range from $\sim$1--4 orders of magnitude smaller than the total abundances of amino acids in the same meteorites. A  general conformity is seen between the pattern given by the total nucleobase abundances and the pattern given by the total amino acid abundances when considering solely the CM meteorites. Total abundances of nucleobases in the CM2 meteorites generally maintain both an increasing order, and a separation of approximately 2 orders of magnitude from the amino acids (with the main exception being the Murchison Smith. meteorite). Total abundances of nucleobases in the CM1 meteorites maintain a decreasing order and a separation of approximately 1--2 orders of magnitude from the amino acids. The patterns given by the total nucleobase and total amino acid abundances disagree in the sections containing the CI and CR meteorites. Though, more meteoritic nucleobase assays need to be performed on these meteorite subclasses to verify that the disagreement in patterns is not the result of a poor sample size. This general conformity between total nucleobases and total amino acids in only the CM meteorites agrees with the above suggestion: that there may be similar reaction mechanisms for nucleobase and amino acid synthesis within the CM meteorite parent bodies, but perhaps there are different reaction and decay rates (e.g. due to different temperatures, hydration) or chemical compositions within the CR and CI meteorite parent bodies.

\subsection{Relative Nucleobase Frequencies}

\begin{figure}[ht!]
\centering
\includegraphics[width=80mm]{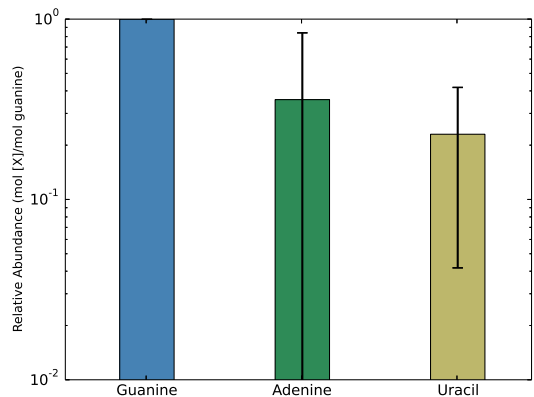}
\caption{Average relative nucleobase abundances to G in CM2 meteorites}
\label{RelativeAbundances}
\end{figure}

Since CM2 meteorites were the most common to host nucleobases and had the highest abundances, we averaged the relative abundances of A and U with respect to G from CM2 meteorites in Figure~\ref{RelativeAbundances}. Here we can see G is the most common nucleobase in carbonaceous chondrites, followed by A at $0.36 \pm0.48$ and then U at $0.23 \pm0.19$. 

\section{Abiotic Nucleobase Synthesis}

We now turn to the question of what chemical reactions might be responsible for nucleobase formation in planetesimals. To pursue this we first consider each well known abiotic nucleobase-synthesizing mechanism, and we filter out the mechanisms that require environments dissimilar to the interior of a planetesimal. We then consider the most frequently discussed reaction pathways for each remaining mechanism, and filter these reactions down to a candidate list based on assumed reactant and catalyst availability within planetesimals.

\subsection{Reaction Mechanisms}

Here we categorize the known types of reactions that synthesize nucleobases abiotically into seven groups. The following list of reaction types is in rough order of decreasing likelihood of occurrence within planetesimals.

Fischer-Tropsch (FT) synthesis uses carbon monoxide, hydrogen and ammonia gases as reactants in the presence of a catalyst (such as nickel-iron alloy or aluminum oxide) to create all five nucleobases \citep{Hayatsu1968,Hayatsu1972,YangOro,Anders1974}. Fischer-Tropsch synthesis is already thought by some to be the mechanism for producing meteoritic organics such as purines, pyrimidines, amino acids and aromatic hydrocarbons \citep{Anders1974,Hayatsu1972}.

Several non-catalytic (NC) chemical reactions have been theorized or proven to synthesize all five nucleobases which only involve a solution being heated, cooled or maintained at room temperature. The reactants in these reactions vary, although HCN is found to be the most common \citep{LaRoweRegnier,HillOrgel2002,SchwartzBakker1989,VoetSchwartz,Yamada1969,Yamada1968,Ferris1978,Wakamatsu1966,OroKimball1962,OroKimball1961}.
HCN is also one of the main reactants in Strecker synthesis, which is proposed to be the reaction that produces amino acids within meteorite parent bodies \citep{Cobb2014,Peltzer1984}.

Various other catalytic (CA) reactions have synthesized A, T, C and U \citep{Kumar2014,Saladino2001,Subbaraman1980,Oro1961b,Bendich1948}. The most common of these reactions uses neat formamide as the sole reactant, which successfully forms nucleobases in the presence of several different catalysts.

Spark discharge experiments first introduced by Stanley L. Miller have produced G and A \citep{Yuasa1984,Miyakawa2000,Miyakawa1999}. These experiments send a high voltage electrical discharge through gaseous mixtures of reactants, e.g., methane, ethane and ammonia or water, carbon monoxide and nitrogen, to produce organics.
  
Ultraviolet irradiation in various solutions, e.g., water, ice and urea, or water, ammonia and pyrimidine, has produced G, C and U \citep{MenorMarin,Nuevo2012,Lemmon1970}. Photo-dehydrogenation, i.e., using ultraviolet light to remove hydrogen from molecules, of 5,6-dihydrouracil and 5,6-dihydrothymine has yielded U and T \citep{ChittendenSchwartz1976,SchwartzChittenden1977}. And finally, electron irradiation in a gaseous mixture of methane, ammonia and water has successfully synthesized A \citep{Ponnamperuma1963}.

This research is on the potential synthesis of nucleobases within planetesimals, thus only the mechanisms listed above that could possibly occur in such environments are of interest. Since the rocky, solid and liquid water-filled interior of a planetesimal \citep{TravisSchubert} is likely not an environment where electron beams, ultraviolet light  or electric discharges would occur, we only consider FT, NC and CA reactions as likely producers of the nucleobases measured in carbonaceous chondrites.

\subsection{Reaction Pathways}

To compile a thorough list of FT, NC and CA reactions that have been employed or suggested to produce nucleobases, we use a combination of Internet browsing with search criteria relevant to nucleobase synthesis (e.g. ``abiotic adenine synthesis'', ``Fischer-Tropsch synthesis of nucleobases'', etc.), and reading through citations from nucleobase synthesis studies.

The 60 most discussed FT, NC and CA nucleobase-synthesizing reactions in scientific literature are formulated in Table~\ref{Schemes}, along with their experimental nucleobase yields. All of the reactions in Table~\ref{Schemes} have successfully synthesized a nucleobase in the laboratory with the exception of the 5 reaction pathways from \citet{LaRoweRegnier}. These reaction pathways have been studied for their thermodynamic potential in hydrothermal environments, and have been proven to be favourable from a theoretical standpoint. If a nucleobase synthesis study suggests a chemical equation for the synthesis of their nucleobase(s), the chemical equation from the study is copied into the reaction column of Table~\ref{Schemes}. For studies that do not suggest a chemical equation, a simple
\begin{equation} \label{reaction}
  reactants \rightarrow nucleobase
\end{equation}
scheme is used. In the case of FT reactions, liquid water is also added as a potential product due to the fact that it generally forms along with nucleobases in the laboratory experiments \citep{Hayatsu1981}. For the deamination of C into U (reaction 32), ammonia is added as a potential product in order to perfectly balance the reaction.

For catalytic reactions, the chemical equation includes one or more catalyst(s) written above the reaction arrow. In the case of multiple catalysts, a '+' is used to signify 'and,' a '$\|$' is used to signify 'or' and a '+$\|$' is used to signify 'and/or.'

The reactions in Table~\ref{Schemes} are  firstly grouped by product nucleobase; in order of decreasing number of reactions. Then each nucleobase group is sub-grouped by reaction type (starting with Fischer-Tropsch synthesis). Finally the reaction type sub-groups are ordered by increasing total molecular mass of reactants (ignoring coefficients).

\section{Results: Candidate Reaction Pathways within Planetesimals}

In order for a reaction pathway from Table~\ref{Schemes} to be considered as a candidate reaction within planetesimals, the reactants and (if applicable) catalyst(s) must be present within the planetesimal environment. Since comets are thought to contain material that would have also been incorporated into planetesimals during the latter's formation \citep{SchulteShock2004,Alexander2011}, we consider all reaction schemes with no reactant cometary abundances unlikely to occur in planetesimals. In the case of catalytic reactions, the catalyst(s) must also be found in meteorites in order for the reaction scheme to be considered as a candidate.

Many spectroscopic molecular surveys and one $\emph{in situ}$ molecular analysis (done with mass spectrometry), have been carried out on various comets \citep{MummaCharnley,Liu2007,Bockelee2004,Crovisier2004,EhrenfreundCharnley,Bockelee2000}. From these surveys only eight of the reactants in Table~\ref{Schemes} have been measured: H$_{2}$, H$_{2}$O, CH$_{2}$O, NH$_{3}$, CO, HCN, Cyanoacetylene and Formamide. All reaction pathways with reactants not from the list of eight above are disregarded. Deuterated hydrogen has only been found in comets in the form of HDO, DCN, HDCO, NH$_{2}$D, HDS, CH$_{3}$OD and CH$_{2}$DOH \citep{Crovisier2004}, though since D$_{2}$ and ND$_{3}$ are not substantially different from H$_{2}$ and NH$_{3}$, we do not immediately disregard the deuterated FT reactions from \citet{Hayatsu1972} as candidates.

The temperature limit for the interior of carbonaceous chondrite parent bodies based on the theoretical temperature for which amino acid synthesis in planetesimals generally shuts off is $\sim$300$^{\circ}$C \citep{CobbPudritzPearce}. In thermal evolution simulations of the interiors of carbonaceous chondrite parent bodies, \citet{TravisSchubert} found temperatures reached a maximum of 180$^{\circ}$C. \citet{McSween2002} had similar results, with their smallest-radius carbonaceous chondrite parent bodies reaching a maximum interior temperature of 227$^{\circ}$C. From a study that classifies temperature ranges within the parent bodies of the various carbonaceous chondrite subclasses and petrologic types, \citet{Sephton2002} assigns a temperature limit of 400$^{\circ}$C for the parent bodies of the carbonaceous chondrites in which nucleobases have been found. Furthermore, based on nucleobase decomposition experiments from \citet{Levy1998}, all five nucleobases decompose in seconds at temperatures above 400$^{\circ}$C in an aqueous environment.

Since the FT synthesis of A, G, U and T in \citet{Hayatsu1972} were only successful when temperatures reached as high as 700$^{\circ}$C, we disregard these reactions as candidates. This temperature is much higher than those within CM, CR and CI carbonaceous chondrite parent bodies, and much too high to sustain nucleobases in aqueous environments \citep{TravisSchubert}.

\citet{Jarosewich} and \citet{Wiik} have done complete chemical analyses of several carbonaceous chondrites. From these two surveys, only Al$_{2}$O$_{3}$, SiO$_{2}$, Ni and Fe (satisfying NiFe) of the above catalysts has been found. This allows us to disregard schemes 23, 25, 50 and 60, as the catalysts for these reactions were unlikely to have been available within carbonaceous chondrite parent bodies.

The remaining reaction schemes complete our list of candidate reaction pathways for the synthesis of nucleobases within planetesimals. This list is presented in Table~\ref{Candidates} and the corresponding reactions are indicated by asterisks in Table~\ref{Schemes}. In total there are 7 candidate reactions for A, 2 for U, 3 for C, 2 for G and 1 for T.

Although some of the reactions in Table~\ref{Schemes} that did not make the candidate list have experimental yields that are indeed high, these pathways require reactants that may not be present in planetesimals. For this study, we limit our candidates to the reactions for which there is some evidence of the reactants being present within meteorite parent bodies.

\section{Discussion}

\begin{deluxetable}{lll}
\tablecolumns{3}
\tablecaption{Candidate reaction pathways for the production of nucleobases within meteorite parent bodies. The reaction numbers pertain to those in Table~\ref{Schemes}.\label{Candidates}}
 \tablehead{Nucleobase&Reaction Number&Type}
 \startdata
Adenine & 1 & FT\\
 & 3 & NC\\
 & 4 & NC\\
 & 6 & NC\\
 & 7 & NC\\
 & 8 & NC\\
 & 24 & CA\\
 & & \\
Uracil & 29 & NC \\
 & 32 & NC \\
  & & \\
Cytosine & 43 & FT\\
 & 44 & NC\\
 & 49 & CA\\
  & & \\
Guanine & 51 & FT \\
 & 54 & NC\\
  & & \\
Thymine & 58 & NC \\[-3mm]
 \enddata
 \tablenotetext{}{NC: Non-catalytic; CA: Catalytic; FT: Fischer-Tropsch.}
\end{deluxetable}

Because nucleobases, nucleobase catabolic intermediates and nucleobases analogs are present in carbonaceous chondrites, these molecules would have likely also been present on the pregenetic, meteorite-accreting Earth. C and T have yet to be found in meteorites. A plausible astrophysical source of C and T could arise from ultraviolet radiation-driven chemical reactions on icy particles in space \citep{Nuevo2014}. This would supply the C needed for the onset of the RNA world that likely preceded DNA-based life \citep{Neveu2013}, as well as the T needed for the later appearance of the DNA, RNA and protein world.

It is currently unknown why C or T have not been measured in carbonaceous chondrites. Photodissociation is unlikely to be the main cause as \citet{Pilling2008} have found that G, A, C, T and U have similar survival rates when exposed to vacuum UV photons at energies ranging $\sim$4--30 eV. Deamination could explain the lack of C in meteorites, since C deaminates to U in aqueous solutions \citep{RobertsonMiller1995,GarrettTsau,Ferris1968}, and meteorite parent bodies are thought to remain aqueous on the order of millions of years \citep{TravisSchubert}.

Perhaps the most likely explanation of the deficit in C and T is a purely chemical one. C and T are known to readily oxidize into 5-hydroxyhydantoin and 5-methyl-5-hydroxyhydantoin respectively \citep{Redrejo}, which could explain the lack of both C and T in meteorites. Although 5-hydroxyhydantoin and 5-methyl-5-hydroxyhydantoin haven't been measured in carbonaceous chondrites, a few amino acid hydantoins have, e.g., 5-(2-Carboxyethyl)hydantoin \citep{Cooper1995}.

A final possibility for the lack of C and T in meteorites could be that the reactions synthesizing A, G and U in meteorite parent bodies are more thermodynamically favourable than those synthesizing C and T. This would result in a reduced production of C and T with respect to the other nucleobases within planetesimals, which could then be further diminished to negligible amounts from photodissociation, deamination or oxidation. An important target of our future investigations is to address this question, i.e., how much C or T could be synthesized within meteorite parent bodies.

Since the abundances of xanthine and hypoxanthine measured in carbonaceous chondrites are comparable to those of G and A, the former would have been just as available as the latter during the evolution of the genome and the genetic code. Xanthine and hypoxanthine were therefore not selected to have a place in the genome for a reason other than availability. 

What is particularly interesting, is the low abundance of nucleobase analogs that have been measured in carbonaceous chondrites. Nucleobase analogs are structurally similar to nucleobases, and 2,6-diaminopurine, like A, can actually base pair with T or U \citep{Kirnos}. If nucleobase analogs can replace some of the genome nucleobases, but were perhaps only available in very small ($<$10 ppb) abundances on the pregenetic Earth, then it is possible to consider meteoritic abundance as a driving factor in A's incorporation into the genome over these nucleobase analogs.

Total nucleobase abundances in carbonaceous chondrites are consistently lower than amino acids by $\sim$1--4 orders of magnitude. The decay rates of molecules due to hydrolysis have been measured and a useful tool in describing them is Arrhenius plots. These results show that nucleobases decay more rapidly than glycine, alanine, serine, glutamic acid, aspartic acid and phenylalanine at temperatures above 200$^{\circ}$C \citep{Levy1998,Sato,Qian}. Though the amino acids studies were not performed at temperatures lower than 200$^{\circ}$C, by extrapolating the Arrhenius curves from \citet{Sato} and \citet{Qian}, it can be seen that amino acids should be more stable than nucleobases at all aqueous temperatures. Since carbonaceous chondrite parent bodies are thought to have had aqueous interiors for millions of years \citep{TravisSchubert}, the $\sim$1--4 orders of magnitude difference in abundance between amino acids and nucleobases within carbonaceous chondrites could simply be a result of the more rapid hydrolysis decay rate of nucleobases.

It is clear from the number of individual nucleobase reactions in Tables~\ref{Schemes} that A is the most commonly produced nucleobase in the lab. Five out of the seven candidate reactions producing A are NC reactions (all involving HCN), producing a maximum 15 \% yield. G on the other hand has only one theoretical NC reaction candidate, with no laboratory yield. Since G happens to be the most abundant nucleobase within carbonaceous chondrites (Figure~\ref{RelativeAbundances}), and A is the second most abundant, NC reactions are perhaps less likely to be the main reaction mechanism to produce nucleobases within meteorite parent bodies. CA reactions are also not an obvious choice for the main reaction mechanism since they are missing a reaction pathway for the G and U found in meteorites.

The last candidate reaction mechanism to consider is FT synthesis, which produces G, A and C. In experiments by \citet{Hayatsu1968}, A had the highest yield (0.16 \%) followed by G (0.09 \%) and C (0.05 \%). This mechanism may at first seem to be an unlikely candidate since C and T are not found in carbonaceous chondrites. However, it is possible that reaction 32 accounts for the missing C in carbonaceous chondrites, as in this reaction C deaminates into U. Though G is on average the most abundant nucleobase in CM2 meteorites, there are also a few instances where A is more abundant than G in CR2, CR3, CM1 and CM2 meteorites. This increases the plausibility of the FT reaction pathways in \citet{Hayatsu1968} occurring within meteorite parent bodies.

\section{Conclusions}

Unlike amino acids, hydroxy acids, carboxylic acids, hydrocarbons and many other organic molecules, nucleobases have not been found in great (ppm-range) abundances in meteorites.
G, A and U have been found in carbonaceous chondrites in the 1--500 ppb range. No C or T has been measured in any meteorites to date. Nucleobase catabolic intermediates xanthine and hypoxanthine have been measured in similar abundances to nucleobases (3--2300 ppb). Nucleobase analogs purine, 2,6-diaminopurine and 6,8-diaminopurine have been measured in low abundances around 0.2--9 ppb.

CM2 meteorites appear to be the preferred type for nucleobases with an average total nucleobase abundance of $330 \pm250$ ppb. Conversely, CR2 meteorites are the least abundant in nucleobases, with an average total abundance of $16 \pm13$ ppb. These results are dissimilar to amino acid abundances in carbonaceous chondrites, where CR2 and CM2 meteorites both contain the greatest abundances. This may suggest that amino acids and nucleobases could share a reaction mechanism within the parent bodies of CM2 meteorites, but may have differing reaction mechanisms within the parent bodies of CR2 meteorites.

Nucleobases are typically 1--3 orders of magnitude less abundant in carbonaceous chondrites than glycine (which is the most abundant amino acid in carbonaceous chondrites). Nucleobases can be 2 orders of magnitude less abundant to even 1 order of magnitude more abundant in carbonaceous chondrites than aspartic acid (one of the less abundant amino acids in carbonaceous chondrites). The most abundant nucleobase in carbonaceous chondrites is guanine, ranging from 2--515 ppb. Adenine is the 2nd most abundant nucleobase with an average relative abundance to G of $0.36 \pm0.48$ within CM2 meteorites. Uracil is the least abundant nucleobase with an average relative abundance to G of $0.23 \pm0.19$ within CM2 meteorites. The total abundance of nucleobases in carbonaceous chondrites is 1--4 orders of magnitude less than the total abundance of amino acids. 

Of the 60 most discussed FT, NC and CA nucleobase-synthesizing reactions in the literature, 3 FT, 10 NC and 2 CA reactions are proposed as the candidate reactions producing nucleobases within meteorite parent bodies. It appears as if the FT synthesis reaction mechanism best complies with the nucleobases found in carbonaceous chondrites thus far.
\newline

We are indebted to Alyssa Cobb, whose work on meteoritic amino acids provided an important template for our study here. We would also like to thank Dr. Allyson Brady for sharing her superb knowledge on molecular analysis techniques, and Dr. Paul Ayers for his advisement on the molecular stability of C and T. The research of B.K.D.P. was supported by a NSERC CREATE Canadian Astrobiology Training Program Undergraduate Fellowship. R.E.P. is supported by an NSERC Discovery Grant.

\clearpage
\begin{turnpage}
\begin{deluxetable*}{lllll}
\tablecolumns{4}
\tablecaption{Reaction schemes and experimental yields for FT, NC and CA nucleobase synthesis.\label{Schemes}}
 \tablehead{No.&Type&Reaction&Max Yield (\%)&Source(s)}
 \startdata
$\underline{Adenine}$ & \\
1\tablenotemark{*} & FT & CO + H$_{2}$ + NH$_{3}$ $\xrightarrow{NiFe+||Al_{2}O_{3}+||SiO_{2}}$ A + H$_{2}$O & detected & \citet{YangOro};\\
 & & & 0.16 & \citet{Hayatsu1968}\\
2 & FT & CO + D$_{2}$ + ND$_{3}$ $\xrightarrow{NiFe+Al_{2}O_{3}}$ A + H$_{2}$O & ++ & \citet{Hayatsu1972}\\
3\tablenotemark{*} & NC & 5HCN$_{(aq)}$ $\rightarrow$ A$_{(aq)}$ & theoretical & \citet{LaRoweRegnier}\\
4\tablenotemark{*} & NC & HCN + NH$_{3}$ $\rightarrow$ A & detected & \citet{Yamada1969};\\
 & & & 15 & \citet{Wakamatsu1966}\\
5 & NC & Ammonium cyanide $\rightarrow$ A & 0.029 & \citet{Miyakawa2002};\\
 & & & 0.038 & \citet{Levy1999}\\
6\tablenotemark{*} & NC & 5CO + 5NH$_{3}$ $\rightarrow$ A + 5H$_{2}$0 & detected & \citet{Hayatsu1968}\\
7\tablenotemark{*} & NC & HCN + H$_{2}$0 $\rightarrow$ A & 0.04 &\citet{Ferris1978}\\
8\tablenotemark{*} & NC & HCN + NH$_{3}$ + H$_{2}$0 $\rightarrow$ A & 0.05 & \citet{OroKimball1961}\\
9 & NC & Ammonium cyanide + H$_{2}$0 $\rightarrow$ A & detected & \citet{Oro1960}\\
10 & NC & HCN + Glycolonitrile $\rightarrow$ A & 0.06 & \citet{SchwartzBakker1989}\\
11 & NC & HCN + Ammonium formate $\rightarrow$ A & 18 &\citet{HillOrgel2002}\\
12 & NC & 4-aminoimidazole-5-carbonnitrile + HCN $\rightarrow$ A & 24 & \citet{Sanchez1968}\\
13 & NC & 4-aminoimidazole-5-carbonnitrile + Ammonium cyanide $\rightarrow$ A & 11 & \citet{Sanchez1968}\\
14 & NC & 4-aminoimidazole-5-carbonnitrile + Sodium cyanide $\rightarrow$ A & 7 & \citet{Sanchez1968}\\
15 & NC & KCN + Ammonium formate + Formamide $\rightarrow$ A & 4.7 & \citet{Hudson2012}\\
16 & NC & NaCN + HCN + Acetamidine hydrocloride + NH$_{3}$ $\rightarrow$ A & detected & \citet{Yamada1968}\\
17 & NC & NaCN + Ammonium chloride + Formamidine acetate + NH$_{3}$ $\rightarrow$ A & 25 & \citet{Yamada1968}\\
18 & NC & Diaminomaleonitrile + Formamidine acetate + NH$_{3}$ $\rightarrow$ A & 1 & \citet{Sanchez1967}\\
19 & NC & 4-amino-5-cyanoimidazole + Formamidine acetate + H$_{2}$0 $\rightarrow$ A & 3 & \citet{FerrisOrgel1966}\\
20 & NC & 4-Formamido-5-imidazolecarboxamidine + Potassium bicarbonate $\rightarrow$ A & 80 & \citet{Shaw1950}\\
21 & NC & 6-chloropurine + Ammoniacal butanol $\rightarrow$ A & detected & \citet{Bendich1954}\\
22 & NC & Aminomalonitrile $\emph{p}$-toluenesulfonate + NaCN + Formamidine acetate + NH$_{3}$ $\rightarrow$ A & 21 & \citet{Yamada1968}\\
23 & CA & 5HCN + 4NH$_{3}$ $\xrightarrow{NH_{4}OH}$ A + 4NH$_{3}$ & detected & \citet{OroKimball1962};\\
 & & & detected & \citet{Oro1961b}\\
24\tablenotemark{*} & CA & Formamide $\xrightarrow{Al_{2}O_{3}||SiO_{2}||Kaolin||Zeolite}$ A &  0.09 & \citet{Saladino2001}\\
25 & CA & Formamide $\xrightarrow{Zn_{2}[Mo(CN)_{8}]\cdot2H_{2}O}$ A & 0.62 & \citet{Kumar2014}\\
26 & CA & Thioisoguanine sulfate $\xrightarrow{Raney-Nickel}$ A & 21.25 & \citet{Bendich1948}\\
$\underline{Uracil}$ & \\
27 & FT & CO + D$_{2}$ + ND$_{3}$ $\xrightarrow{NiFe+Al_{2}O_{3}}$ U + H$_{2}$O & + & \citet{Hayatsu1972}\\
28 & NC & Ammonium cyanide $\rightarrow$ U & 0.0017 & \citet{Miyakawa2002}\\
29\tablenotemark{*} & NC & 2HCN$_{(aq)}$ + 2CH$_{2}$O$_{(aq)}$ $\rightarrow$ U$_{(aq)}$ + H$_{2(aq)}$ & theoretical & \citet{LaRoweRegnier}\\
30 & NC & HCN + Ammonium hydroxide + H$_{2}$O $\rightarrow$ U & 0.005 & \citet{VoetSchwartz}\\
31 & NC & 2,4-Diaminopyrimidine + H$_{2}$O $\rightarrow$ U & 36 & \citet{Ferris1974}\\
32\tablenotemark{*} & NC & C + H$_{2}$O $\rightarrow$ U + NH$_{3}$ & detected & \citet{RobertsonMiller1995};\\
 & & & detected & \citet{GarrettTsau};\\
 & & & 95 & \citet{Ferris1968}\\
33 & NC & Cyanoacetaldehyde + Urea $\rightarrow$ U & 0.08 & \citet{Nelson2001}\\
34 & NC & Acrylonitrile + Urea + H$_{2}$O $\rightarrow$ U & $<$ 1 & \citet{GrossenbacherKnight} \\
35 & NC & Acrylonitrile + Urea + Ammonium chloride $\rightarrow$ U & $<$ 1 & \citet{Oro1963}\\
36 & NC & Propiolic acid + Urea + Sulfuric acid + H$_{2}$O $\rightarrow$ U & 59 & \citet{HaradaSuzuki}\\
37 & NC & Propiolic acid + Urea + Polyphosphoric acid + H$_{2}$O $\rightarrow$ U & 75 & \citet{HaradaSuzuki}\\
38 & NC & Fumaric acid + Urea + Polyphosphoric acid $\rightarrow$ U & 5 & \citet{TakamotoYamamoto}\\
39 & NC & Maleic acid + Urea + Polyphosphoric acid $\rightarrow$ U & 20 & \citet{TakamotoYamamoto}\\
40 & NC & Malic acid + Urea + Sulfuric acid $\rightarrow$ U & 55 & \citet{DavidsonBaudisch}\\
41 & NC & Malic acid + Urea + Polyphosphoric acid $\rightarrow$ U & 9.3 & \citet{FoxHarada1961}\\
42 & CA & Acetylene dicarboxylic acid + Urea $\xrightarrow{NH_{4}H_{2}PO_{4}||(NH_{4})_{2}HPO_{4}||NH_{4}Cl||H_{3}PO_{4}}$ U & 6.3 & \citet{Subbaraman1980}\\[-3mm]
 \enddata
\end{deluxetable*}
\end{turnpage}
\clearpage
\global\pdfpageattr\expandafter{\the\pdfpageattr/Rotate 90}

\clearpage
\begin{turnpage}
\begin{deluxetable*}{lllll}
\tablecolumns{4}
\tablenum{3}
\tablecaption{$\textit{Continued}$}
 \tablehead{No.&Type&Reaction&Max Yield (\%)&Source(s)}
 \startdata
$\underline{Cytosine}$ & \\
43\tablenotemark{*} & FT & CO + H$_{2}$ + NH$_{3}$ $\xrightarrow{NiFe+||Al_{2}O_{3}+||SiO_{2}}$ C + H$_{2}$O & detected & \citet{YangOro};\\
 & & & 0.05 & \citet{Hayatsu1968}\\
44\tablenotemark{*} & NC & 3HCN$_{(aq)}$ + CH$_{2}$O$_{(aq)}$ $\rightarrow$ C$_{(aq)}$ & theoretical & \citet{LaRoweRegnier}\\
45 & NC & Cyanoacetaldehyde + Urea $\rightarrow$ C & 0.23 & \citet{Nelson2001};\\
 & & & 53 & \citet{RobertsonMiller1995}\\
46 & NC & 2,4-Diaminopyrimidine + H$_{2}$O $\rightarrow$ C & trace & \citet{Ferris1974}\\
47 & NC & Cyanoacetylene + Urea + H$_{2}$O $\rightarrow$ C & 5 & \citet{Ferris1968}\\
48 & NC & Cyanoacetylene + Potassium cyanate + H$_{2}$O $\rightarrow$ C & 19 & \citet{Ferris1968};\\
 & & & 5 &\citet{Sanchez1966}\\
49\tablenotemark{*} & CA & Formamide $\xrightarrow{Al_{2}O_{3}||SiO_{2}||Kaolin||Zeolite}$ C & 0.44 & \citet{Saladino2001}\\
50 & CA & Formamide $\xrightarrow{Zn_{2}[Mo(CN)_{8}]\cdot2H_{2}O}$ C & 0.36 & \citet{Kumar2014}\\
$\underline{Guanine}$ & \\
51\tablenotemark{*} & FT & CO + H$_{2}$ + NH$_{3}$ $\xrightarrow{NiFe+||Al_{2}O_{3}+||SiO_{2}}$ G + H$_{2}$O & detected & \citet{YangOro};\\
 & & & 0.09 & \citet{Hayatsu1968}\\
52 & FT & CO + D$_{2}$ + ND$_{3}$ $\xrightarrow{NiFe+Al_{2}O_{3}}$ G + H$_{2}$O & +++ & \citet{Hayatsu1972}\\
53 & NC & Ammonium cyanide $\rightarrow$ G & 0.0067 & \citet{Miyakawa2002};\\
 & & & 0.0035 & \citet{Levy1999}\\
54\tablenotemark{*} & NC & 5HCN$_{(aq)}$ + H$_{2}$O $\rightarrow$ G$_{(aq)}$ + H$_{2(aq)}$ & theoretical & \citet{LaRoweRegnier}\\
55 & NC & 4-aminoimidazole-5-carboxamide + Cyanogen $\rightarrow$ G & 43 & \citet{Sanchez1968}\\
56 & NC & 4-aminoimidazole-5-carboxamide + Potassium cyanate $\rightarrow$ G \; \; \; \; \; \; \; \; \; \; \; \; \; \; \; \; \; & 20 & \citet{Sanchez1968}\\
$\underline{Thymine}$ & \\
57 & FT & CO + D$_{2}$ + ND$_{3}$ $\xrightarrow{NiFe+Al_{2}O_{3}}$ T + H$_{2}$O & + & \citet{Hayatsu1972}\\
58\tablenotemark{*} & NC & 2HCN$_{(aq)}$ + 3CH$_{2}$O$_{(aq)}$ $\rightarrow$ T$_{(aq)}$ + H$_{2}$O & theoretical & \citet{LaRoweRegnier}\\
59 & NC & U + Paraformaldehyde + Hydrazine $\rightarrow$ T & 0.1 & \citet{StephenSherwood}\\
60 & CA & Formamide $\xrightarrow{Zn_{2}[Mo(CN)_{8}]\cdot2H_{2}O}$ T & 0.03 & \citet{Kumar2014}\\[-3mm]
 \enddata
 \tablenotetext{*}{Candidate reactions referenced in Table~\ref{Candidates}.}
 \tablenotetext{}{NC: Non-catalytic; CA: Catalytic; FT: Fischer-Tropsch.}
 \tablenotetext{}{+++ very abundant; ++ abundant; + less abundant.}
\end{deluxetable*}
\end{turnpage}
\clearpage
\global\pdfpageattr\expandafter{\the\pdfpageattr/Rotate 90}

\end{document}